\documentclass[12pt]{article}
\usepackage{amssymb}

\newcommand{\nc}{\newcommand}
\nc{\fh}{\hat{f}}
\nc{\muh}{\hat{\mu}}
\nc{\nuh}{\hat{\nu}}
\nc{\bib}{\bibitem}
\nc{\al}{\alpha}
\nc{\g}{\gamma}
\nc{\G}{\Gamma}
\nc{\D}{\Delta}
\nc{\eps}{\epsilon}
\nc{\la}{\lambda}
\nc{\La}{\Lambda}
\nc{\var}{\varphi}
\nc{\cg}{{\cal G}}
\nc{\pa}{\partial}
\nc{\nn}{\nonumber \\ }
\nc{\hf}{\frac{1}{2}}  
\nc{\dz}{\frac{dz}{2\pi i}}
\nc{\bin}[2]{\left (\begin{array}{c} {#1}\\ {#2} \end{array}\right )}
\nc{\ben}{\begin{equation}}
\nc{\een}{\end{equation}}
\nc{\bea}{\begin{eqnarray}}
\nc{\eea}{\end{eqnarray}}
\nc{\bra}[1]{\langle {#1}|}
\nc{\ket}[1]{|{#1}\rangle}

\def\min{{\rm min}}

\begin{document}

\topmargin -5mm
\oddsidemargin 5mm

\begin{titlepage}
\setcounter{page}{0}

\vspace{8mm}
\begin{center}
{\huge Jordan cells in logarithmic limits}\\[.3cm]
{\huge of conformal field theory}

\vspace{15mm}
{\Large J{\o}rgen Rasmussen}\\[.3cm] 
{\em Department of Mathematics and Statistics, University of Melbourne}\\ 
{\em Parkville, Victoria 3010, Australia}\\[.3cm]
j.rasmussen@ms.unimelb.edu.au

\end{center}

\vspace{10mm}
\centerline{{\bf{Abstract}}}
\vskip.4cm
\noindent
It is discussed how a limiting procedure of conformal field theories
may result in logarithmic conformal field theories with Jordan
cells of arbitrary rank. This extends our work on rank-two
Jordan cells. We also consider the limits of certain three-point functions
and find that they are compatible with known results.
The general construction is illustrated by logarithmic
limits of minimal models in conformal field theory.
Characters of quasi-rational representations are found to emerge
as the limits of the associated irreducible Virasoro characters. 
\\[.5cm]
{\bf Keywords:} Logarithmic conformal field theory,
minimal models, Virasoro characters. 
\\[.1cm]
{\bf PACS number:} 11.25.Hf
\end{titlepage}
\newpage
\renewcommand{\thefootnote}{\arabic{footnote}}
\setcounter{footnote}{0}

\section{Introduction}

Results on logarithmic conformal field theory (CFT) 
(see \cite{Flo,Gab,Nic} for recent reviews of logarithmic CFT, and
\cite{FMS} for a survey on ordinary CFT)
are often worked out for Jordan cells of rank two followed by a statement
that the results may be extended to higher-rank Jordan cells.
This is not always ensured a priori, however.
Our construction of rank-two Jordan cells in
logarithmic limits of CFTs \cite{Ras}
provides an example where the extension is quite
non-trivial. Its resolution is discussed in the present paper.
Certain indecomposable representations involving Jordan cells 
of rank three are analyzed in \cite{EF}.
Jordan cells of infinite rank have been introduced in \cite{infrank},
while affine Jordan cells relevant to logarithmic extensions of
Wess-Zumino-Witten models have been constructed in \cite{affine}.

The idea of the present construction is to consider a sequence of 
conformal models labelled by an integer $n$, with focus
on a multiplet of primary fields in each conformal model appearing
in the sequence. Each multiplet consists of $r$ fields where $r$
is the rank of the Jordan cell we wish to construct.
To get a firmer grip on this, we introduce sequences of
primary fields and organize the former in equivalence classes. 
For finite $n$, the fields in a given multiplet must all
have different conformal
weights, while the weights of the associated sequences 
converge to the same (finite)
conformal weight, $\D$, as $n$ approaches infinity.
A Jordan-cell structure of rank $r$ emerges if one considers a
particular linear and (for finite $n$) invertible map
of the multiplet of fields (or of the associated multiplet of sequences)
into a new multiplet of $r$ fields. Since the
original fields have different conformal weights, the new
fields do not all have well-defined conformal weights.
In the limit $n\rightarrow\infty$, the linear map is singular 
and thus not invertible (thereby mimicking the In\"on\"u-Wigner or
Saletan contractions known from the theory of Lie algebras), 
while the new multiplet of fields
make up a rank-$r$ Jordan cell of conformal weight $\D$.

The two-point functions of the new fields are discussed in
generality. Particular three-point functions are also
considered and found to be compatible with known results
on logarithmic CFT.

To further support the idea that a logarithmic CFT
may appear as the limit of a sequence of minimal models,
we study the limits of the corresponding sequences of irreducible
Virasoro characters. We find that the characters of the so-called
quasi-rational representations naturally appear as the
limiting characters. These quasi-rational representations
are believed to play an important role in logarithmic CFT,
see \cite{GK,PRZ}, for example. Here we discuss certain aspects
of the structure of these representations, in particular in regards
to singular vectors and irreducibility.
We also address how indecomposable
representations based on Jordan cells can emerge as
a result of the limiting procedure, and 
discuss the associated characters.

Section 2 concerns the general construction of 
higher-rank Jordan cells in logarithmic CFT obtained as a limiting procedure
of ordinary CFTs. The construction is illustrated by logarithmic limits
of minimal models in Section 3 which also contains a discussion
of quasi-rational representations as limits of irreducible representations. 
Concluding remarks may be found in Section 4.

\section{Logarithmic limits}

A Jordan cell of rank $r=\rho+1$ consists of one primary field, $\Psi_0$, and
$\rho$ logarithmic partners, $\Psi_1,...,\Psi_\rho$, satisfying \cite{RAK}
\ben
 T(z)\Psi_{j}(w)\ =\ 
 \frac{\D\Psi_{j}(w)+(1-\delta_{j,0})\Psi_{j-1}(w)}{(z-w)^2}
 +\frac{\pa_w\Psi_{j}(w)}{z-w},\ \ \ \ \ \ \ j=0,1,...,\rho
\label{r}
\een
where $\Psi_{-1}\equiv0$. Their two-point functions read
\bea
 \langle\Psi_i(z)\Psi_j(w)\rangle&=&0,\ \ \ \ \ \ \ \ i+j<\rho\nn
 \langle\Psi_i(z)\Psi_\rho(w)\rangle&=&\frac{\sum_{m=0}^i\frac{(-2)^m}{m!}
   A_{i-m}\left(\ln(z-w)\right)^m}{(z-w)^{2\D}}\nn
 \langle\Psi_i(z)\Psi_j(w)\rangle&=&
  \langle\Psi_{i+j-\rho}(z)\Psi_\rho(w)\rangle\nn
 &=&\frac{\sum_{m=0}^{i+j-\rho}\frac{(-2)^m}{m!}
   A_{i+j-\rho-m}\left(\ln(z-w)\right)^m}{(z-w)^{2\D}},
  \ \ \ \ \ \ \ i+j\geq\rho
\label{twopoint}
\eea
Our goal is to construct such a system 
in the limit of a sequence of ordinary CFTs. 
We shall work with real structure constants 
$A_j$, $j=0,...,\rho$, and assume that $A_0>0$.

Let us consider a sequence of conformal models $M_n$, 
$n\in\mathbb{Z}_>$, with central charges, $c_n$, converging to the finite value
\ben
 \lim_{n\rightarrow\infty}c_n\ =\ c
\label{c}
\een
It is assumed that $M_n$ contains a multiplet of $r$ primary fields, 
$\var_{0;n},...,\var_{\rho;n}$, that is,
\ben
 T_n(z)\var_{j;n}(w)\ =\ \frac{\D_{j;n}\var_{j;n}(w)}{(z-w)^2}
  +\frac{\pa_w\var_{j;n}(w)}{z-w}
\label{Tvar}
\een
where $T_n$ is the Virasoro generator in $M_n$. 
The conformal weights for given $n$ are all different and may be written 
\ben
 \D_{j;n}\ =\ \D+a_{j;n},\ \ \ \ \ \ \ 0\leq j\leq\rho
\label{D}
\een
with $\D$ independent of $n$. The multiplets are organized or ordered
so that
\ben
 a_{0;n}>a_{1;n}>...>a_{\rho;n}
\label{aaa}
\een
Despite these differences, the corresponding sequences
of conformal weights all approach $\D$, as we require that
\ben
 \lim_{n\rightarrow\infty}a_{j;n}\ =\ 0,\ \ \ \ \ \ \ 0\leq j\leq\rho
\label{a}
\een
We are thus considering multiplets (labelled by $j$, $0\leq j\leq\rho$)
of sequences of primary fields such as
$(\var_{j;1},\var_{j;2},...)$,
where the element $\var_{j;n}$ belongs to $M_n$.
The associated two-point functions are of the form
\ben
 \langle\var_{i;n}(z)\var_{j;n}(w)\rangle\ =\ 
   \frac{\delta_{ij}C_{j;n}}{(z-w)^{2\D_{j;n}}}
\label{twopointvar}
\een
where, for simplicity, 
one may normalize the fields so that the non-vanishing
structure constants, $C_{j;n}$, are {\em independent} of $n$.
This is the preferred choice in \cite{Ras} on rank-two Jordan cells 
but is not necessary and is {\em not adopted} here.
For simplicity, we shall assume, though, that $C_{j;n}>0$.

Consider now the linear and invertible map
\ben
 \Psi_{i;n}\ =\ \sum_{j=0}^\rho F_{i,j;n} \var_{j;n}
\label{lin}
\een
governed by a sequence of invertible $r\times r$ matrices, $F_n$.
We shall be interested in the limit 
\ben
 \Psi_j:=\ \lim_{n\rightarrow\infty}\Psi_{j;n}
\label{Psin}
\een
and look for a sequence of maps (\ref{lin}) 
that would result in a Jordan-cell 
structure of rank $r$ as $n$ approaches infinity. That is, we must work out
\bea
 T(z)\Psi_j(w)&=&\lim_{n\rightarrow\infty}\left\{T_n(z)\Psi_{j;n}(w)\right\}\nn
 \langle\Psi_i(z)\Psi_j(w)\rangle&=&\lim_{n\rightarrow\infty}\left\{
   \langle\Psi_{i;n}(z)\Psi_{j;n}(w)\rangle\right\}
\label{limit}
\eea
and extract conditions on $F_n$ from a comparison with 
(\ref{r}) and (\ref{twopoint}), and ultimately try to find a suitable
sequence of such maps.

To this end we shall denote the diagonal matrix 
diag$[a_{0;n},...,a_{\rho;n}]$ by $a_n$ and introduce the
off-diagonal $r\times r$ matrix
\ben
 P\ =\ \left(\begin{array}{ccccccc} 0&0&0&...&0&0&0\\
   1&0&0&...&0&0&0\\   0&1&0&...&0&0&0\\  \vdots&&&\vdots&&&\vdots\\
   \vdots&&&\vdots&&&\vdots\\
    0&0&0&...&1&0&0\\  0&0&0&...&0&1&0  \end{array}\right)
\label{P}
\een
where $P_{i,j}=\delta_{i,j+1}$.
The conditions arising from (\ref{r}) then read
\ben
 \lim_{n\rightarrow\infty}\left\{F_na_nF^{-1}_n\right\}\ =\ P
\label{FaFP}
\een
while the conditions from the two-point functions (\ref{twopoint}) are
\ben
 \lim_{n\rightarrow\infty}\left\{\sum_{\ell=0}^\rho C_{\ell;n} F_{i,\ell;n}
   F_{j,\ell;n}a_{\ell;n}^m\right\}\ =\ 
 \left\{\begin{array}{ll} A_{i+j-\rho-m},
  \ \ \ \ \ &
     0\leq m\leq i+j-\rho\\ \\  0,&{\rm otherwise} \end{array}\right.
\label{CFF}
\een
The solution for $F_n$ is not unique. The following is the simplest one
we have found:
\bea
 F_{j,k;n}&=&0,\hspace{7cm} 0\leq j<k\leq\rho\nn
 F_{k,k;n}&=&\sqrt{(-1)^k}\times
  \sqrt{\al_{k;n}\prod_{\ell=1}^{\rho-k}(a_{k;n}-a_{k+\ell;n})},
  \ \ \ \ \ \ \ \  0\leq k\leq\rho\nn
 F_{j,k;n}&=&\frac{F_{k,k;n}}{\prod_{\ell=1}^{j-k}(a_{k;n}-a_{k+\ell;n})},
  \hspace{3.7cm}0\leq k<j\leq\rho
\label{Fal}
\eea
where
\ben
 \al_{k;n}\ = \ 
   \frac{1}{C_{k;n}}\sum_{j=0}^k\frac{(-1)^jA_j}{
  \prod_{\ell=1}^{k-j}(a_{j+\ell-1;n}-a_{k;n})},
    \ \ \ \ \ \ 0\leq k\leq\rho
\label{all}
\een
The factor $\sqrt{(-1)^k}$ corresponds to multiplication by 
$i$ when $k$ is odd or by 1 when $k$ is even. Up to this
explicit and potentially imaginary factor,
all entries of $F_n$ are non-negative and real for large $n$. 
This follows from (\ref{aaa}) and that $C_{k;n}$ and $A_0$
have been assumed positive. Allowing these structure
constants to be negative (or even complex) would merely
affect the transparency of the notation in (\ref{Fal}).

To prove that (\ref{Fal}) indeed respects the conditions (\ref{FaFP}) and
(\ref{CFF}), we first factorize $F_n$ as
\ben
 F_n\ =\ {\cal A}_n\times {\rm diag}[F_{0,0;n}, F_{1,1;n},...,F_{\rho,\rho;n}]
\label{FAF}
\een
where the $r\times r$ matrix ${\cal A}_n$ is given by
\bea
 {\cal A}_{i,j;n}&=&0,\hspace{4cm}0\leq i<j\leq\rho\nn
 {\cal A}_{j,j;n}&=&1,\hspace{4cm}0\leq j\leq\rho\nn
 {\cal A}_{i,j;n}&=&\frac{1}{\prod_{k=j+1}^i(a_{j;n}-a_{k;n})},\ \ \ \ \ 
   0\leq j<i\leq\rho\
\label{A}
\eea
Its inverse is found to be
\bea
  {\cal A}_{i,j;n}^{-1}&=&0,\hspace{4cm}0\leq i<j\leq\rho\nn
 {\cal A}_{j,j;n}^{-1}&=&1,\hspace{4cm}0\leq j\leq\rho\nn
 {\cal A}_{i,j;n}^{-1}&=&\frac{(-1)^{i-j}}{
  \prod_{k=j}^{i-1}(a_{k;n}-a_{i;n})},\ \ \ \ \ \ \ \ 
   0\leq j<i\leq\rho\
\label{A-}
\eea
By inserting (\ref{FAF}) into the condition (\ref{FaFP}), the latter is seen 
to become independent of the various structure constants as it reduces to
\ben
 \lim_{n\rightarrow\infty}\left\{{\cal A}_na_n{\cal A}^{-1}_n\right\}\ =\ P
\label{AaAP}
\een
The non-trivial part reads
\bea
 \delta_{i,j+1}&=&\lim_{n\rightarrow\infty}\left\{\frac{\sum_{k=j}^i(-1)^{k-j}
   a_{k;n}\prod_{j\leq u<v\leq i; u,v\neq k}
   (a_{u;n}-a_{v;n})}{\prod_{j\leq u<v\leq i}(a_{u;n}-a_{v;n})}\right\}
\label{AP}
\eea
Likewise, the left side of (\ref{CFF}) becomes
\bea
 \lim_{n\rightarrow\infty}\left\{\sum_{\ell=0}^\rho C_{\ell;n} F_{i,\ell;n}
   F_{j,\ell;n}a_{\ell;n}^m\right\}
  &=&\lim_{n\rightarrow\infty}\left\{
    \sum_{\ell=0}^{\min(i,j)}(-1)^\ell\sum_{t=0}^{\ell}\frac{(-1)^tA_t}{
    \prod_{s=1}^{\ell-t}(a_{s+t-1;n}-a_{\ell;n})}\right.\nn
  & &\times\left. \frac{a_{\ell;n}^m\prod_{k=1}^{\rho-\ell}(a_{\ell;n}-a_{\ell+k;n})
   }{\prod_{u=1}^{i-\ell}(a_{\ell;n}-a_{\ell+u;n})\times
    \prod_{v=1}^{j-\ell}(a_{\ell;n}-a_{\ell+v;n})}\right\}
\label{CFFA}
\eea
which is obviously symmetric in $i$ and $j$ (as (\ref{CFF}) is). 
We may thus choose to consider $j\leq i$ in
which case the non-trivial part of the condition (\ref{CFF}) reduces to
\bea
&& \delta_{i+j,\rho+m+t}\\
 &=&\lim_{n\rightarrow\infty}\left\{\sum_{\ell=t}^{j}
  (-1)^{\ell-t}\frac{a_{\ell;n}^m\prod_{k=1}^{\rho-i}(a_{\ell;n}-a_{i+k;n})}{
   \prod_{v=1}^{j-\ell}(a_{\ell;n}-a_{\ell+v;n})\times
    \prod_{s=1}^{\ell-t}(a_{s+t-1;n}-a_{\ell;n})}\right\}\nn
  &=&\lim_{n\rightarrow\infty}\left\{\frac{\sum_{\ell=t}^j(-1)^{\ell-t}a_{\ell;n}^m
   \prod_{k=1}^{\rho-i}(a_{\ell;n}-a_{i+k;n})\times
  \prod_{t\leq u<v\leq j;u,v\neq\ell}
   (a_{u;n}-a_{v;n})}{\prod_{t\leq u<v\leq j}(a_{u;n}-a_{v;n})}\right\}
\label{deltam}
\eea

To finally verify (\ref{AP}) and (\ref{deltam}),
one may employ the following simple observation:
Let $Q_N(x_1,...,x_M)$ denote a homogeneous polynomial
in the $M$ variables $x_1,...,x_M$. Its degree is bounded by $N$ as
deg$(Q_N)\leq N$, and $Q_N\equiv0$ for $N<0$.
If, for every pair $(i,j)$ with $1\leq i<j\leq M$, $Q_N=0$ when
$x_i=x_j$, we may conclude that
\ben
 Q_N(x_1,...,x_M)\ =\ \prod_{1\leq i<j\leq M}(x_i-x_j)\times
  Q_{N-\frac{1}{2}M(M-1)}(x_1,...,x_M)
\label{Q}
\een
Here $Q_{N-\frac{1}{2}M(M-1)}$ may be zero. This is obviously
the case if $N<\frac{1}{2}M(M-1)$.
That the Kronecker delta functions in (\ref{AP}) and (\ref{deltam}) appear
without prefactors, is easily checked.

In conclusion, we have found that Jordan cells of arbitrary rank
may emerge in limits of certain sequences of ordinary CFTs.

The solution for rank-two Jordan cells was found in \cite{Ras}.
With the notation used here, it is given by
\bea
 F_{0,0;n}&=&\sqrt{\frac{A_0(a_{0;n}-a_{1;n})}{C_{0;n}}},\ \ \ \ \ \ \ \ \ \ \ 
   F_{0,1;n}\ =\ 0\nn
 F_{1,0;n}&=&\sqrt{\frac{A_0}{C_{0;n}(a_{0;n}-a_{1;n})}},\ \ \ \ \ \ \ \ \
  F_{1,1;n}\ =\ 
  i\sqrt{\frac{A_0-A_1(a_{0;n}-a_{1;n})}{C_{1;n}(a_{0;n}-a_{1;n})}}
\label{two}
\eea
For rank three, the solution reads
\bea
 F_{0,0;n}&=&\sqrt{\frac{A_0(a_{0;n}-a_{1;n})(a_{0;n}-a_{2;n})}{C_{0;n}}}\nn
 F_{1,0;n}&=&\sqrt{\frac{A_0(a_{0;n}-a_{2;n})}{C_{0;n}(a_{0;n}-a_{1;n})}}\nn
 F_{2,0;n}&=&\sqrt{\frac{A_0}{C_{0;n}(a_{0;n}-a_{1;n})(a_{0;n}-a_{2;n})}}\nn
 F_{1,1;n}&=&i\sqrt{\frac{\{A_0-A_1(a_{0;n}-a_{1;n})\}
  (a_{1;n}-a_{2;n})}{C_{1;n}(a_{0;n}-a_{1;n})}}\nn
 F_{2,1;n}&=&i\sqrt{\frac{A_0-A_1(a_{0;n}-a_{1;n})}{C_{1;n}
  (a_{0;n}-a_{1;n})(a_{1;n}-a_{2;n})}}\nn
 F_{2,2;n}&=&\sqrt{\frac{A_0-\{A_1-A_2(a_{1;n}-a_{2;n})\}(a_{0;n}-a_{2;n})}{
   C_{2;n}(a_{0;n}-a_{2;n})(a_{1;n}-a_{2;n})}}\nn
 F_{0,1;n}&=&F_{0,2;n}\ =\ F_{1,2;n}\ =\ 0
\label{three}
\eea

\subsection{Three-point functions}

The objective here is to indicate that our construction is
compatible with known results on
three-point functions in logarithmic CFT.
We only consider the coupling of three identical 
logarithmic fields in a Jordan cell of rank two,
but find that the results provide further
evidence to the sensibility of our construction.

Let $\Psi$ denote the only logarithmic field in the rank-two Jordan cell
resulting in the limiting procedure outlined above.
The three-point function of our interest thus reads
\bea
&& \langle\Psi(z_1)\Psi(z_2)\Psi(z_3)\rangle\nn
&&\ \ \ \ \ \ \ \ \ \  =\ \lim_{n\rightarrow\infty}\left\{
   \langle\left(F_{1,0;n}\var_{0;n}(z_1)+F_{1,1;n}\var_{1;n}(z_1)\right)\times
   \left(F_{1,0;n}\var_{0;n}(z_2)+F_{1,1;n}\var_{1;n}(z_2)\right)\right.\nn
 &&\ \ \ \ \ \ \ \ \ \ \ \ \ \ \ \ \ \ \ \  
  \left.\times\left(F_{1,0;n}\var_{0;n}(z_3)+F_{1,1;n}\var_{1;n}(z_3)\right)\rangle\right\}
\label{threelim}
\eea 
Using that the structure constants of the three-point functions
\bea
 &&\langle\var_{j_1;n}(z_1)\var_{j_2;n}(z_2)\var_{j_3;n}(z_3)\rangle\nn
 &=& \frac{C_{j_1,j_2,j_3;n}}{(z_1-z_2)^{\D_{1;n}+\D_{2;n}-\D_{3;n}}
 (z_2-z_3)^{-\D_{1;n}+\D_{2;n}+\D_{3;n}}(z_1-z_3)^{\D_{1;n}-\D_{2;n}+\D_{3;n}}}
\label{threevar}
\eea
are symmetric for each given $n$, and introducing the common abbreviation
$z_{ij}=z_i-z_j$, we find
\bea
 &&\langle\Psi(z_1)\Psi(z_2)\Psi(z_3)\rangle
 \ =\ \frac{1}{(z_{12}z_{23}z_{13})^{\D}}\nn
 &\times&\lim_{n\rightarrow\infty}\{ 
  \left[F_{1,0;n}^3C_{0,0,0;n}+3F_{1,0;n}^2F_{1,1;n}C_{1,0,0;n}
   +3F_{1,0;n}F_{1,1;n}^2C_{1,1,0;n}+F_{1,1;n}^3C_{1,1,1;n}\right]\nn
 &&+\left[-a_{0;n}F_{1,0;n}^3C_{0,0,0;n}-(2a_{0;n}+a_{1;n})F_{1,0;n}^2F_{1,1;n}C_{1,0,0;n}
  \right.\nn
 &&\ \ \ \left.-(a_{0;n}+2a_{1;n})F_{1,0;n}F_{1,1;n}^2C_{1,1,0;n}
  -a_{1;n}F_{1,1;n}^3C_{1,1,1;n}\right]
  \ln(z_{12}z_{23}z_{13})\nn
 &&+\left[\frac{1}{2}a_{0;n}^2F_{1,0;n}^3C_{0,0,0;n}
  +(2a_{0;n}^2-2a_{0;n}a_{1;n}+\frac{3}{2}a_{1;n}^2)
  F_{1,0;n}^2F_{1,1;n}C_{1,0,0;n}
  \right.\nn
 &&\ \ \ \left.(\frac{3}{2}a_{0;n}^2-2a_{0;n}a_{1;n}+2a_{1;n}^2)F_{1,0;n}F_{1,1;n}^2C_{1,1,0;n}
  +\frac{1}{2}a_{1;n}^2F_{1,1;n}^3C_{1,1,1;n}\right]\nn
 &&\ \ \ \times\left(\ln^2(z_{12})+\ln^2(z_{23})+\ln^2(z_{13})\right)\nn
&&+\left[a_0^2F_{1,0;n}^3C_{0,0,0;n}
  +(4a_{0;n}a_{1;n}-a_{1;n}^2)
  F_{1,0;n}^2F_{1,1;n}C_{1,0,0;n}
  \right.\nn
 &&\ \ \ \left.(-a_{0;n}^2+4a_{0;n}a_{1;n})F_{1,0;n}F_{1,1;n}^2C_{1,1,0;n}
  +a_{1;n}^2F_{1,1;n}^3C_{1,1,1;n}\right]\nn
 &&\ \ \ \times\left(\ln(z_{12})\ln(z_{23})+\ln(z_{23})\ln(z_{13})
  +\ln(z_{13})\ln(z_{12})\right)+\dots\}
\label{threeln}
\eea
This should be compared to the known results for three-point functions
in logarithmic CFT (see \cite{logWard} and references therein),
that is, $\langle\Psi(z_1)\Psi(z_2)\Psi(z_3)\rangle$ should be of the form
\bea
 \langle\Psi(z_1)\Psi(z_2)\Psi(z_3)\rangle
 &=&\frac{1}{(z_{12}z_{23}z_{13})^{\D}}\{B_0+B_1\ln(z_{12}z_{23}z_{13})\nn
 &&+B_2\left(\ln^2(z_{12})+\ln^2(z_{23})+\ln^2(z_{13})\right)\nn
 &&+B_3\left(\ln(z_{12})\ln(z_{23})+\ln(z_{23})\ln(z_{13})
  +\ln(z_{13})\ln(z_{12})\right)\}
\label{B}
\eea
To facilitate the comparison, we choose the normalization
\ben
 C_{0;n}\ =\ C_{1;n}\ =\ \frac{A_0}{a_0-a_1}
\label{normC}
\een 
in which case (\ref{two}) gives
\bea
 F_{0,0;n}^3&=&1\nn
 F_{1,0;n}^2F_{1,1;n}&=&i\left(1-\frac{1}{2}\frac{A_1}{A_0}
  (a_{0;n}-a_{1;n})-\frac{1}{8}\frac{A_1^2}{A_0^2}(a_{0;n}-a_{1;n})^2
   +{\cal O}((a_{0;n}-a_{1;n})^3)\right)\nn
 F_{1,0;n}F_{1,1;n}^2&=&-\left(1-\frac{A_1}{A_0}(a_{0;n}-a_{1;n})\right)\nn 
 F_{1,1;n}^3&=&-i\left(1-\frac{3}{2}\frac{A_1}{A_0}
  (a_{0;n}-a_{1;n})+\frac{3}{8}\frac{A_1^2}{A_0^2}(a_{0;n}-a_{1;n})^2
   +{\cal O}((a_{0;n}-a_{1;n})^3)\right)\ \ \ 
\label{twored}
\eea
Let us also introduce the following expansions of the
three-point structure constants (which of course may depend on $n$)
\bea
 C_{0,0,0;n}&=&\frac{\al_0}{(a_{0;n}-a_{1;n})^2}+
  \frac{\beta_0}{a_{0;n}-a_{1;n}}+\gamma_0+{\cal O}(a_{0;n}-a_{1;n})\nn
 C_{1,0,0;n}&=&\frac{\al_1}{(a_{0;n}-a_{1;n})^2}+
  \frac{\beta_1}{a_{0;n}-a_{1;n}}+\gamma_1+{\cal O}(a_{0;n}-a_{1;n})\nn
 C_{1,1,0;n}&=&\frac{\al_2}{(a_{0;n}-a_{1;n})^2}+
  \frac{\beta_2}{a_{0;n}-a_{1;n}}+\gamma_2+{\cal O}(a_{0;n}-a_{1;n})\nn
 C_{1,1,1;n}&=&\frac{\al_3}{(a_{0;n}-a_{1;n})^2}+
  \frac{\beta_3}{a_{0;n}-a_{1;n}}+\gamma_3+{\cal O}(a_{0;n}-a_{1;n})
\label{abc}
\eea
Based on these, the comparison of (\ref{threeln}) with (\ref{B}) yields
\ben
 B_3\ =\ -2B_2
\label{BB}
\een
and
\bea
 -\al_0-i\al_1\ =\ i\al_1-\al_2\ =\ \al_2+i\al_3&=&B_2\nn
 \beta_0+2i\beta_1-\beta_2\ =\ -i\beta_1+2\beta_2+i\beta_3
  &=&-B_1+\frac{A_1}{A_0}B_2\nn
 \gamma_0+3i\gamma_1-3\gamma_2-i\gamma_3&=&B_0+
  \frac{3}{2}\frac{A_1}{A_0}B_1-\frac{3}{4}\frac{A_1^2}{A_0^2}B_2
\label{abcB}
\eea
As non-trivial results, we thus have that $B_3$ and $B_2$ are
related as in (\ref{BB}), and that the easily solved linear system
(\ref{abcB}) consists of three decoupled systems.
Imposing the consistency condition (\ref{BB}) on (\ref{B})
corresponds to restricting to the special three-point functions discussed 
in \cite{FloCluster} (see also \cite{GG}), cf. \cite{logWard}.
We conclude that our construction appears to yield sensible results
for three-point functions.

\section{Minimal models}

Here we illustrate the general construction above by 
considering limits of minimal models. The minimal model
${\cal M}(p,p')$ is characterized by the coprime integers $p$
and $p'$ which may be chosen to satisfy $p>p'>1$.
The central charge is given by
\ben
 c\ =\ 1-6\frac{(p-p')^2}{pp'}
\label{cmm}
\een
whereas the primary fields, $\phi_{r,s}$, have conformal weights given by
\ben
 \D_{r,s}\ =\ \frac{(rp-sp')^2-(p-p')^2}{4pp'}, \ \ \ \ \ \ \ \ \ 
   1\leq r<p', \ \ \ 1\leq s< p
\label{Dmm}
\een
The bounds on $r$ and $s$ define the Kac table of admissible
primary fields. With the identification 
\ben
 \phi_{r,s}\ =\ \phi_{p'-r,p-s}
\label{iden}
\een
there are $(p-1)(p'-1)/2$ distinct primary fields in the model.
These models are unitary provided $p=p'+1$.

We now follow our recent work on rank-two Jordan cells \cite{Ras}.
For each positive integer $k$, we thus consider the sequence of minimal
models ${\cal M}(kn+1,n)$, $n\geq2$. The central charges
and conformal weights are given by
\bea
 c^{(k;n)}&=&1-6\frac{((k-1)n+1)^2}{n(kn+1)}\nn
  &=&1-6\frac{(k-1)^2}{k}-6\frac{(k^2-1)}{k^2n}+{\cal O}(1/n^2)\nn
 \D_{r,s}^{(k;n)}&=&\frac{((kn+1)r-ns)^2-((k-1)n+1)^2}{4n(kn+1)}\nn
  &=&\frac{(kr-s)^2-(k-1)^2}{4k}+\frac{k^2(r^2-1)-(s^2-1)}{4k^2n}
  +{\cal O}(1/n^2)
\label{kn}
\eea
with limits
\bea
 c^{(k)}&=&\lim_{n\rightarrow\infty}c^{(k;n)}\ =\ 1-6\frac{(k-1)^2}{k}\nn
 \D_{r,s}^{(k)}&=&\lim_{n\rightarrow\infty}\D_{r,s}^{(k;n)}\ =\ 
  \frac{(kr-s)^2-(k-1)^2}{4k},\ \ \ \ \ \ \ r,s\in\mathbb{Z}_>
\label{k}
\eea
As discussed in \cite{Ras}, these are seen to correspond to the similar 
values in the (non-minimal) model ${\cal M}(k,1)$ with {\em extended} Kac table
in which $r$ and $s$ are unbounded from above.
In particular, the spectrum of the extended model ${\cal M}(1,1)$ with central
charge $c^{(1)}=1$ is thereby related to 
the limit of the sequence of {\em unitary} minimal models ${\cal M}(n+1,n)$. 
A different approach to rank-two Jordan cells and
logarithmic CFT based on minimal models may be found in \cite{Flo-9605}.

There is a natural embedding of the Kac table associated to 
${\cal M}(kn_1+1,n_1)$ into the Kac table associated to
${\cal M}(kn_2+1,n_2)$ if $n_1\leq n_2$, mapping 
$\phi_{r,s}^{(k,n_1)}$ to $\phi_{r,s}^{(k,n_2)}$.
It is noted, however, that the conformal weights and representations
in general will be altered. Our point here is that if
$(r,s)$ is admissible for $n_0$, it will be admissible
for all $n\geq n_0$. We thus have a natural notion of
sequences of primary fields: $(\phi_{r,s}^{(k,n_0)},
\phi_{r,s}^{(k,n_0+1)},...)$.
The parameter $n_0$ is essentially immaterial since we are
concerned with the properties of the sequences as $n\rightarrow\infty$.
We therefore denote such a sequence simply as 
$\Upsilon_{r,s}^{(k)}$.

These sequences may be organized in equivalence classes, where
$\Upsilon_{r,s}^{(k)}$ and $\Upsilon_{u,v}^{(k)}$ 
are said to be equivalent if they approach the same conformal
weight. According to \cite{Ras}, this presupposes that
\ben
  I:\ \ (u,v)\ =\ (r+q,s+kq),\ \ \ \ \ \ r,s,u,v\in\mathbb{Z}_>,\ \ q\in\mathbb{Z}
\label{I}
\een
or
\ben
 II:\ \ (u,v)\ =\ (-r+q,-s+kq),\ \ \ \ \ \ r,s,u,v\in\mathbb{Z}_>,\ \ q\in\mathbb{Z}
\label{II}
\een
In either case, the approached conformal weight is $\D_{r,s}^{(k)}$ given in
(\ref{k}). The equivalence becomes trivial (i.e.,
$\Upsilon_{r,s}^{(k)}=\Upsilon_{u,v}^{(k)}$) if $q=0$ in case $I$
or if $q=2r$ and $s=kr$ in case $II$.
With the notation (cf. (\ref{D}))
\ben
 \D_{r,s}^{(k;n)}\ =\ \D_{r,s}^{(k)}+a_{r,s;n}^{(k)}
\label{DDa}
\een
we have
\bea
 a_{r,s;n}^{(k)}-a_{r+q,s+qk;n}^{(k)}&=&\frac{q(2n(s-rk)-2r-q)}{4n(kn+1)}\nn
 &=&\frac{q(s-rk)}{2kn}-\frac{q(2s+qk)}{4k^2n^2}+{\cal O}(1/n^3)\nn
 a_{r,s;n}^{(k)}-a_{-r+q,-s+qk;n}^{(k)}&=&
   \frac{q(2n(rk-s)+2r-q)}{4n(kn+1)}\nn
 &=&\frac{q(rk-s)}{2kn}+\frac{q(2s-qk)}{4k^2n^2}+{\cal O}(1/n^3)
\label{DD}
\eea

According to the general prescription outlined in the previous section,
one can now construct a Jordan cell of {\em arbitrary rank} for each
conformal dimension $\D_{r,s}^{(k)}$ 
in the spectrum of the extended ${\cal M}(k,1)$ (\ref{k}), i.e., for each pair $(r,s)$
with $r,s\geq1$. To avoid confusion with
the conventional labelling of primary fields employed here, we shall
denote the rank by $\rho+1$.
For $rk<s$ one may consider the following multiplet of sequences
\ben
 \Upsilon_{r+q_0,s+q_0k}^{(k)},\ \Upsilon_{r+q_1,s+q_1k}^{(k)},\ ...\ ,\ 
  \Upsilon_{r+q_\rho,s+q_\rho k}^{(k)},\ \ \ \ \ \ \ \ \ 0\leq q_0<q_1<...<q_\rho
\label{rk<s}
\een
It is thus ordered to comply with (\ref{aaa}). For $rk\geq s$ one may
consider the ordered multiplet
 \ben
 \Upsilon_{r+q_0,s+q_0k}^{(k)},\ \Upsilon_{r+q_1,s+q_1k}^{(k)},\ ...\ ,\ 
  \Upsilon_{r+q_\rho,s+q_\rho k}^{(k)},\ \ \ \ \ \ \ \ \ q_0>q_1>...>q_\rho\geq0
\label{rk>s}
\een
These multiplets are obviously not unique as one may choose to work
with case $II$ (\ref{II}) instead or even combinations of the two cases.
This general construction of Jordan cells of rank $\rho+1$
applies to all positive integer $k$.

\subsection{Characters and quasi-rational representations}

It is recalled (see \cite{FMS}, for example) that the Virasoro character of
the irreducible representation corresponding to $(r,s)$ in
the Kac table associated to ${\cal M}(p,p')$ may be written
\ben
 \chi_{r,s}(q)\ =\ K_{pr-p's}(q)-K_{pr+p's}(q)
\label{chirs}
\een
where
\ben
 K_{\lambda}(q)\ =\ \frac{1}{\eta(q)}
  \sum_{m\in\mathbb{Z}}q^{(\lambda+2mpp')^2/4pp'}
\een
with the Dedekind $\eta$ function given by
\ben
 \eta(q)\ =\ q^{1/24}\prod_{m=1}^\infty(1-q^m)
\een
For given $(r,s)$, we introduce the sequence of such characters where
a single character is defined in ${\cal M}(kn+1,n)$ for each $n\geq n_0$.
As in the case of sequences of primary fields, $n_0$ is 
essentially immaterial since we are concerned with the 
behaviour  as $n\rightarrow\infty$.
To study this, we examine
\bea
 K_{(kn+1)r-ns}(q)&=&\frac{1}{\eta(q)}
  \sum_{m\in\mathbb{Z}}q^{((kn+1)r-ns+2mn(kn+1))^2/4n(kn+1)}\nn
 &=&\frac{1}{\eta(q)} \sum_{m\in\mathbb{Z}}
   q^{\frac{((kn+1)r-ns)^2}{4n(kn+1)}+m((kn+1)r-ns+mn(kn+1))}
\label{Kn}
\eea
and it follows that, as $n\rightarrow\infty$, the only finite power of $q$ occurs
for $m=0$.
The sequence of characters $\chi_{r,s}(q)$ in ${\cal M}(kn+1,n)$ thus
approaches
\bea
 \chi_{r,s}(q)&\rightarrow&\frac{1}{\eta(q)}\lim_{n\rightarrow\infty}
  \left(q^{\frac{((kn+1)r-ns)^2}{4n(kn+1)}}
   -q^{\frac{((kn+1)r+ns)^2}{4n(kn+1)}}\right)\nn
 &=&\frac{1}{\eta(q)}
  \left(q^{\frac{(kr-s)^2}{4k}}-q^{\frac{(kr+s)^2}{4k}}\right)
\label{chilim}
\eea
This is recognized as the character of the so-called quasi-rational
representation $Q_{r,s}^{(k,1)}$ (see (\ref{Qrs}) and (\ref{chiQ}) below)
\ben
 \chi_{r,s}^{(k)}(q)\ =\ \frac{q^{(1-c^{(k)})/24}}{\eta(q)}q^{\D_{r,s}^{(k)}}
  \left(1-q^{rs}\right)
\label{chiqr}
\een
associated to ${\cal M}(k,1)$. As already indicated, 
these representations and their characters
exist for {\em all} positive pairs $(r,s)$ corresponding to an {\em extended}
Kac table as opposed to the {\em ordinary} Kac table, cf. (\ref{Dmm}). 

These considerations can be extended straightforwardly to other 
sequences than our main example ${\cal M}(kn+1,n)$. 
As pointed out to us by A. Nichols
and discussed in \cite{Ras},
we could consider a sequence of the form ${\cal M}(pp'n+1,p'^2n)$.
In this case, we have
\bea
 c^{(p,p')}&=&\lim_{n\rightarrow\infty}c^{(p,p';n)}\ =\ 
  \lim_{n\rightarrow\infty}\left(1-6\frac{(pp'n+1-p'^2n)^2}{(pp'n+1)p'^2n}\right)\nn
  &=&1-6\frac{(p-p')^2}{pp'}\nn
 \D_{r,s}^{(p,p')}&=&\lim_{n\rightarrow\infty}\D_{r,s}^{(p,p';n)}\ =\
   \lim_{n\rightarrow\infty}
   \frac{(r(pp'n+1)-sp'^2n)^2-(pp'n+1-p'^2n)^2}{4(pp'n+1)p'^2n}\nn
 &=&\frac{(rp-sp')^2-(p-p')^2}{4pp'}
\eea
These limits correspond to the central charge and the extended
Kac table of the model ${\cal M}(p,p')$. 
The limits of the associated sequences of characters are 
found to be
\bea
 \chi_{r,s}^{(p,p')}(q)&=&\lim_{n\rightarrow\infty}\chi_{r,s}^{(p,p';n)}(q)\ =\
 \lim_{n\rightarrow\infty}\left(K_{(pp'n+1)r-p'^2ns}(q)
   -K_{(pp'n+1)r+p'^2ns}(q)\right)   \nn
 &=&\frac{q^{(1-c^{(p,p')})/24}}{\eta(q)}q^{\D_{r,s}^{(p,p')}}
  \left(1-q^{rs}\right)
\label{chirspp}
\eea
corresponding to the quasi-rational representations $Q_{r,s}^{(p,p')}$
labelled by the extended Kac table of ${\cal M}(p,p')$.

For every pair of coprime positive integers $(p,p')$ and every pair of positive
integers $(r,s)$, the highest-weight Verma module $V_{r,s}^{(p,p')}$ exists. 
It contains a singular vector at level $rs$ from which the submodule
denoted $V_{r,-s}^{(p,p')}$ is generated. This notation is justified by the
simple relation $\D_{r,-s}^{(p,p')}=\D_{r,s}^{(p,p')}+rs$. The quotient module
\ben
 Q_{r,s}^{(p,p')}\ =\ V_{r,s}^{(p,p')}/V_{r,-s}^{(p,p')}
\label{Qrs}
\een
is typically not irreducible but reducible, see below. 
Due to its properties in regards to fusion 
\cite{Nahm,GK,PRZ}, it is often referred to as the {\em quasi-rational}
representation $Q_{r,s}^{(p,p')}$. 
As already announced, its character $\chi(Q_{r,s}^{(p,p')})$ is given 
by the expression in (\ref{chirspp}) since
\ben
 \chi(Q_{r,s}^{(p,p')})\ =\ \frac{q^{(1-c^{(p,p')})/24}}{\eta(q)}
  \Big(q^{\D_{r,s}^{(p,p')}}-q^{\D_{r,-s}^{(p,p')}}\Big)\ =\ 
  \frac{q^{(1-c^{(p,p')})/24}}{\eta(q)}q^{\D_{r,s}^{(p,p')}}\left(1-q^{rs}\right)
\label{chiQ}
\een

Let us comment on the fate of singular vectors and the emergence of 
these quasi-rational representations $Q_{r,s}^{(p,p')}$.
In the minimal model ${\cal M}(\hat{p},\hat{p}')$, the highest-weight Verma module 
$V_{r,s}^{(\hat{p},\hat{p}')}$ with highest weight $\D_{r,s}^{(\hat{p},\hat{p}')}$ is reducible
as it contains two distinct singular vectors at levels $rs$ and
$(\hat{p}'-r)(\hat{p}-s)$, respectively.
The quotient module obtained by dividing out the proper submodules generated from
these two vectors is the associated {\em irreducible} highest-weight module
with highest weight $\D_{r,s}^{(\hat{p},\hat{p}')}$. Now, in the sequences under consideration,
the level $(\hat{p}'-r)(\hat{p}-s)$ of the second singular vector increases as a polynomial
function of $n$ (since $(\hat{p},\hat{p}')=(pp'n+1,p'^2n)$, for example). 
In the limit $n\rightarrow\infty$, the vector in question thus appears
with infinite weight and is discarded along with all other vectors with
infinite weight. The surviving singular vector appears at unchanged level
$rs$. The quotient module obtained by dividing out the submodule generated
from this singular vector 
is the aforementioned quasi-rational quotient module or representation 
$Q_{r,s}^{(p,p')}$ (which may nevertheless abstain from being irreducible, see below). 
The limit of irreducible representations
with Kac labels $(r,s)$ in the sequence of minimal models ${\cal M}(pp'n+1,p'^2n)$, 
for example, thus corresponds to the quasi-rational representation $Q_{r,s}^{(p,p')}$
with extended Kac labels $(r,s)$ associated to ${\cal M}(p,p')$.

This analysis indicates that quasi-rational representations are natural
objects in the model constructed as the limit of certain sequences of minimal models.
This is in accordance with known results
on logarithmic CFT where these quasi-rational representations seem to play
an important role, see \cite{GK,PRZ}, for example.

To illustrate how these considerations of limits of characters are related
to the construction of Jordan cells above, we focus on the sequence
${\cal M}(2n+1,n)$, that is, ${\cal M}(kn+1,n)$ with $k=2$.
The spectrum of the resulting model is thus described in terms of
the extended Kac table of ${\cal M}(2,1)$. Let us consider the 
two equivalent sequences $\Upsilon_{1,1}^{(2)}$ and $\Upsilon_{1,3}^{(2)}$,
cf. (\ref{II}).
Individually, they give rise to the quasi-rational characters
\bea
 \chi_{1,1}^{(2,1)}(q)&=&\frac{q^{1/8}}{\eta(q)}(1-q)
  \ =\ q^{1/12}(1+q^2+q^3+2q^4+2q^5+...)\nn
 \chi_{1,3}^{(2,1)}(q)&=&\frac{q^{1/8}}{\eta(q)}(1-q^3)
  \ =\ q^{1/12}(1+q+2q^2+2q^3+4q^4+5q^5+...)
\label{1113}
\eea
If combined to form a rank-two Jordan cell, that is, $\var_{0;n}=\phi_{1,1}^{(2n+1,n)}$
and $\var_{1;n}=\phi_{1,3}^{(2n+1,n)}$, on the other hand, the result is
quite different. We first note that $a_{0;n}=a_{1,1;n}^{(2)}=0$ and
$a_{1;n}=a_{1,3;n}^{(2)}=-1/2n+{\cal O}(1/n^2)$ thus satisfying (\ref{aaa}).
For finite $n$, the energy levels of excitations in $\var_{0;n}$ will in general not
differ by integers from the energy levels in $\var_{1;n}$ since these differences
are $|\D_{0;n}-\D_{1;n}|$ mod an integer.
As $n\rightarrow\infty$, however, the off-integer parts of these differences vanish.
Also for finite $n$, the action of the Virasoro modes on the two representations
is diagonal in the sense that it does not mix the states in the two modules.
It is built in by construction, though, that in the limit $n\rightarrow\infty$,
the action of the Virasoro modes is {\em non-diagonal}, cf. (\ref{r}).
This means that the resulting module is {\em indecomposable}.
One may view it as an indecomposable combination of the
two quasi-rational representations $Q_{1,1}^{(2,1)}$ and 
$Q_{1,3}^{(2,1)}$ where the 
diagonal parts of the Virasoro action yield these 
modules separately while the indecomposable structure is governed by the
off-diagonal part.

Since the character of this indecomposable
module is insensitive or blind to
the off-diagonal structure of the Jordan cell, it
is given by the {\em sum} of the two characters appearing in
(\ref{1113}). Denoting this character by $\chi_{(1,1),(1,3)}^{(2,1)}(q)$,
we see that is is given by
\ben
 \chi_{(1,1),(1,3)}^{(2,1)}(q)\ =\ \frac{q^{1/8}}{\eta(q)}(2-q-q^3)
  \ =\  q^{1/12}(2+q+3q^2+3q^3+6q^4+7q^5+...)
\een
Possibly up to the finer details of the indecomposable structure,
this indecomposable representation has
already appeared in the literature. 
In \cite{GK}, it is denoted ${\cal R}_{1,1}$ while
it is denoted $(1,1)\oplus_i(1,3)$ in \cite{PRZ}.
Most of the comments above about its emergence as the limit
of a combination of two irreducible representations apply 
generally to our limiting procedure.

Let us also comment on the fact that quasi-rational representations 
may not be irreducible, even though they can arise as limits of irreducible
representations. To appreciate this, we initially consider the 
Verma module $V_{1,3}^{(4,3)}$ in the minimal model ${\cal M}(4,3)$.
It has two proper submodules starting at levels $rs=3$ and $(p'-r)(p-s)=2$,
respectively. The singular vector from which the latter submodule is generated
is
\ben
 \ket{\eta_{1,3}^{(4,3)}}\ =\ \Big(L_{-2}-\frac{3}{2(2\D_{1,3}^{(4,3)}+1)}L_{-1}^2\Big)
  \ket{\D_{1,3}^{(4,3)}}
\label{eta13}
\een
The quasi-rational module $Q_{1,3}^{(4,3)}$ is not irreducible due
to this singular vector.
Now, in the sequence ${\cal M}(12n+1,9n)$, we are considering 
$V_{1,3}^{(12n+1,9n)}$ with proper submodules starting at levels $3$
and $(9n-1)(12n-2)$, respectively. This means, in particular, that there
is no singular vector at level 2. Vectors in $V_{1,3}^{(12n+1,9n)}$ somehow
corresponding to $\ket{\eta_{1,3}^{(4,3)}}$ are
\bea
 \ket{\eta_{1,3}^{(12n+1,9n)}}&=&\Big(L_{-2}-\frac{3}{2(2\D_{1,3}^{(4,3)}+1)}L_{-1}^2\Big)
  \ket{\D_{1,3}^{(12n+1,9n)}}\nn
 \ket{\eta_{1,3}^{(12n+1,9n)}}_{alt}&=&\Big(L_{-2}-\frac{3}{2(2\D_{1,3}^{(12n+1,9n)}+1)}L_{-1}^2\Big)
  \ket{\D_{1,3}^{(12n+1,9n)}}
\label{eta13n}
\eea
though, as already stated, these are not singular. In the limit $n\rightarrow\infty$,
on the other hand, both of them approach $\ket{\eta_{1,3}^{(4,3)}}$ in $V_{1,3}^{(4,3)}$
thereby rendering the module arising in the limit $n\rightarrow\infty$, namely
$Q_{1,3}^{(4,3)}$, {\em reducible}.

\section{Conclusion}

We have discussed how Jordan cells of arbitrary rank may be
constructed in a limiting procedure of ordinary CFTs.
Our construction is quite general and has
been illustrated by (an infinite family of) sequences of minimal
models.
It may also be extended to $N=1$ superconformal field theory.
This is demonstrated explicitly in the rank-two case in our recent work \cite{Ras}.
A somewhat related construction of rank-two Jordan cells
in CFT based on (graded) parafermions is discussed in \cite{BS,Rabat}.  

It is emphasized that the present work is focused
on the mere construction of Jordan cells and the emergence and structure
of quasi-rational representations. Questions regarding the resulting
models being well-defined logarithmic CFTs will be addressed elsewhere.
It is stressed in this context that the only three-point
function we have analyzed is (\ref{threelim}) for rank two and that the
only thing we have verified is that, 
to the degree of our analysis, it is {\em compatible} with
known results in logarithmic CFT. 
Nevertheless, this is already a non-trivial test of our construction.
A more thorough examination of three-point functions
is clearly desirable, in particular in the case of minimal models, 
though beyond the scope of the present work.
It is also stressed that we have only worked with the
particular solution (\ref{two}). A further study may reveal
that a more general solution is required. Even so, we
find that the current analysis adds substantial new evidence to the
suggestion \cite{Ras} that a logarithmic
CFT may result as the limit of a sequence of ordinary CFTs.

It is also emphasized that we are {\em not} claiming that a logarithmic
CFT {\em must} emerge as the limit of a sequence of CFTs. Instead,
we are arguing that a logarithmic CFT {\em may} appear.
As far as the characters go, on the other hand, we have
found that characters of quasi-rational representations {\em always} emerge as
the limit of {\em certain} sequences of irreducible
Virasoro characters. It is obvious, though, that not all
sequences result in characters of quasi-rational representations.
\vskip.5cm
\noindent{\em Acknowledgements}
\vskip.1cm
\noindent  The author thanks M. Flohr, P. Mathieu and P. Pearce for comments.

\end{document}